\documentclass[article,superscriptaddress,twocolumn,amsmath,amssymb,floatfix]{revtex4}
\usepackage{graphicx,psfig}
\usepackage{dcolumn}
\usepackage{bm}
\begin{document}

\title{Self--organized network evolution coupled to extremal dynamics}

\author{Diego Garlaschelli}
\affiliation{Dipartimento di Fisica, Universit\`a di Siena, Via Roma 56, 53100 Siena, Italy.}
\author{Andrea Capocci}

\affiliation{Centro Studi e Ricerche e Museo della Fisica,
  ``E. Fermi'', Compendio Viminale, Roma, Italy.}

\affiliation{Dipartimento di Informatica e Sistemistica,
via Salaria 113, 00185 Roma, Italy.}
\author{Guido Caldarelli}

\affiliation{CNR-INFM Centro SMC, Dip. di Fisica, Universit\`a 
``La Sapienza'', Piazzale A. Moro 5, 00185 Roma, Italy.}

\affiliation{Centro Studi e Ricerche e Museo della Fisica,
  ``E. Fermi'', Compendio Viminale, Roma, Italy.}
\affiliation{
Linkalab, Center for the Study of Complex Networks, Sardegna (Italy).}

\begin{abstract}
The interplay between topology and dynamics in complex networks is a fundamental but poorly explored problem. Here we study this phenomenon on a coupled model describing extremal dynamics on a network which is in turn shaped by the dynamical variable itself. 
Each vertex is assigned a fitness, and the vertex with minimum fitness and its neighbours are updated as in the Bak--Sneppen  
model. Simultaneously, the fitness values determine the connection probability as in the fitness network model. We solve the model analytically and show that the system self--organizes to a nontrivial state which differs from what is obtained when the two processes are decoupled. A power--law decay of dynamical and topological quantities above a threshold emerges spontaneously, as well as a feedback between different dynamical regimes and the underlying network's correlation and percolation properties. Our model is a prototype for more general mechanisms exploring such unexpected effects in networks.
\end{abstract}
\maketitle
The properties of dynamical processes defined on complex networks display a strong dependence on the topology \cite{guidosbook,cosinbook,AB01,siam}. 
On the other hand, there is growing empirical evidence \cite{mywtw,shares,duygu} that  many networks are in turn shaped by some variable associated to each vertex, an aspect captured by the `fitness' or `hidden--variable' model \cite{fitness,soderberg}.
Until now, these two facets of the same problem have been treated as separate, by considering on one hand dynamical processes on static networks \cite{guidosbook,siam}, and on the other hand network formation mechanisms driven by quenched variables \cite{fitness,soderberg,BB01,pastor,servedio}. 
This may be perhaps justified for short time scales. However, in the long--term evolution it is crucial to understand the effects that these mechanisms have on each other, without \emph{ad hoc} specifications of any fixed structure either in the topology or in the dynamical variables. Remarkably, the interplay of dynamics and topology can drive the network to a self--organized state that cannot be inferred by studying the two evolutionary processes as decoupled.

Here we explore explicitly the possibility that the network supports a dynamical process which in turn shapes its topology, with a continuous feedback between dynamics and structure. Models where both dynamical and topological properties are continuously updated have been considered \cite{webworld,jain,ginestra,holyst,zanette,plos}. In these cases, however, the rewiring of links is not completely driven by the dynamical variables. 
By contrast, our main interest here is the description of a self--organized process where the dynamical variable fully acts also as the `hidden variable' shaping network topology explicitly, as in the fitness model.  
Due to the increased complexity of the problem, in this paper we choose the simplest possible dynamical rule for the hidden variable. We focus on the extremal dynamics defined in the Bak--Sneppen (BS) model \cite{BS}, a traditional model of self--organized criticality (SOC) \cite{BTW87} inspired by biological evolution. 
Since the outcomes of this model on a wide range of fitness--independent networks are well known \cite{BS,BSd,BSrn,BSsw,BSsf,BSkim,BSkahng}, it is straightforward to understand what are the novel effects originating uniquely by the interplay with the fitness--driven topological evolution we consider here.

\section*{Coupling the Bak--Sneppen model and the Fitness network model}
In the traditional Bak--Sneppen model \cite{BS} defined on a generic graph \cite{BSd,BSrn,BSsw,BSsf,BSkim,BSkahng}, each of the $N$ vertices is regarded as a biological species having a fitness value $x_i$, initially drawn from a uniform distribution between $0$ and $1$. 
At each timestep the species with lowest fitness and all its neighbours undergo a mutation, and their fitness values are drawn anew from the same uniform distribution. The process is iterated, and eventually the system reaches a stationary state, characterized by a step--like fitness distribution, uniform above a threshold $\tau$. This behaviour is observed on regular lattices \cite{BS,BSd}, random graphs \cite{BSrn}, small--world \cite{BSsw} and scale--free \cite{BSsf,BSkim,BSkahng} networks, the only dependence on the particular topology being in the value of $\tau$\cite{BS,BSd,BSrn,BSsw,BSsf,BSkim,BSkahng}. In particular, $\tau$ vanishes for scale--free degree distributions with diverging second moment\cite{BSsf,BSkim,BSkahng}.

Here we couple this dynamical rule with the fitness model assumption \cite{fitness} that the network is formed by drawing a link between any two vertices $i$ and $j$ with fitness--dependent probability $f(x_i,x_j)$, thus introducing an intrinsic feedback between dynamics and topology. 
In this way, whenever the fitness $x_i$ of a species $i$ is updated to $x'_i$, the links from $i$ to all the other vertices $j$ are drawn anew with probability $f(x'_i,x_j)$. 
Besides the updates described above, one could also define additional and arbitrary \emph{link updating events}. In other words, in addition to the `natural' link update occurring between a mutating species and all other species, other link updates may happen between any two vertices $i$ and $j$ at generic and arbitrarily distributed timesteps. When this occurs, the link between $i$ and $j$ is updated and drawn anew with probability $f(x_i,x_j)$ where $x_i$ and $x_j$ are the current fitness values of $i$ and $j$, even if the latter are not involved in a mutation event. 
Remarkably, it is possible to show (see Supplementary Information) that the introduction of link updating events leaves the system in the same stationary state as if they were absent. Therefore our model is very general in this respect, and allows for rearrangements of ecological interactions on shorter timescales than those generated by mutations. In particular, the stationary state is the same if the \emph{whole} network is updated at each timestep. In this case storing the information on the adjacency matrix among species is unnecessary, and we shall exploit this property to achieve fast and very large numerical simulations of the model. 

As we show below, the coupling between structure and dynamics leads to unexpected results that cannot be traced back to any of the two processes taken as separate. 
Moreover, another important advantage is that the main limitations of the two models disappear when they are coupled together. 
A fundamental problem in the BS model on static graphs is that, after a mutation, the new species always inherits exactly all the links of the previous one. This is hard to justify, since it is precisely the structure of ecological connections among species which is believed to be both the origin and the outcome of macroevolution \cite{webworld}. Here the fitness--driven link updating  overcomes this problem. Similarly, the static fitness model requires the specification of an \emph{ad hoc} fitness distribution that never changes. By contrast, 
here the fitness distribution self--organizes spontaneously to a stationary probability density, removing the need of arbitrary specifications. 
As we discuss below, a proper interpretation of the fitness also allows us to remove the remaining arbitrariness in the choice of $f(x_i,x_j)$. However, in order to keep our approach as general as possible, we first study the model analytically for a generic form of $f$, and focus on particular choices only later.

The analytical solution of the model for an arbitrary linking function $f(x,y)$ can be obtained by focusing on the master equation for the fitness distribution $\rho(x)$ at the stationary state (see Supplementary Information). We find that the analytical expression for $\rho(x)$ is 
\begin{equation}
\rho(x)=\left\{\begin{array}{ll}
(\tau N)^{-1} &x<\tau\\
\displaystyle{\frac{1}{N\int_0^\tau f(x,m)dm}}
&x>\tau
\end{array}\right.
\label{eq_rho2}
\end{equation}
where $\tau$ is a threshold value determined through the normalization condition $\int_0^1\rho(x)dx$, which reads
\begin{equation}
\int_\tau^1\displaystyle{\frac{dx}{\int_0^\tau f(x,m)dm}}=N-1
\label{eq_norm}
\end{equation}
In the infinite size limit $N\to\infty$, the distribution $q(m)$ of the minimum fitness value $m\equiv x_{min}$ is uniform between $0$ and $\tau$, while all other values (except possibly a vanishing fraction) are above $\tau$ (see Supplementary Information). In other words, $q(m)=\Theta(\tau-m)/\tau$. 
This characterizes the stationary state completely. 
Once $\rho(x)$ is known, all the expected topological quantities can be determined as in the static fitness model \cite{fitness,pastor,servedio}.
For instance, the average degree of a vertex with fitness $x$ is given by
\begin{equation}
k(x)=N\int_0^1 f(x,y)\rho(y)dy
\label{eq_kx}
\end{equation}
and the inverse function $x(k)$ can be used to obtain the cumulative degree distribution as 
\begin{equation}
P_>(k)\equiv \int_k^{k(1)}P(k')dk'=\rho_>[x(k)]
\label{eq_pk}
\end{equation} 
where $\rho_>(x)\equiv\int_x^1 \rho (x')dx'$ is the cumulative fitness distribution. 
Note that if $\tau$ is nonzero the fitness distribution preserves the discontinuous behaviour displayed on static networks \cite{BS,BSd,BSrn,BSsw,BSsf,BSkim,BSkahng}. However, here we find the novel feature that $\rho(x)$ is in general not uniform for $x>\tau$. 
This unexpected result holds for any nontrivial choice of $f(x,y)$, and hence for any topology. Therefore the effect is not due to the topology itself, but to the interplay between the topological evolution and the dynamical process entangled with it.
Remarkably, this feedback alone determines the self--organization of the system from a random structure to a complex network with nontrivial dynamical and topological properties.

\section*{Fitness--independent random graphs}
The above analytical solution holds for any form of $f(x,y)$. Now we consider possible choices of this function. 
First note that the null choice is $f(x,y)=p$, the network being a random graph. It is nonetheless an instructive simple case, and we briefly discuss it. Moreover, this choice is asymptotically equivalent to the random neighbour variant \cite{BSrn} of the BS model, the average degree of each vertex being $d=p(N-1)\approx pN$ (we drop terms of order $1/N$ from now on). 
Our analytical results read
\begin{equation}
\rho(x)=\left\{\begin{array}{ll}
(\tau N)^{-1} &x<\tau\\
(p\tau N)^{-1}
&x>\tau
\end{array}\right.
\label{eq_rhop}
\end{equation}
and, depending on how $p$ scales with $N$, eq.(\ref{eq_norm}) implies 
\begin{equation}
\tau=\frac{1}{1+pN}\to
\left\{\begin{array}{lll}
1&pN\to 0\\
(1+d)^{-1}&pN=d\\
0&pN\to \infty
\end{array}\right.
\label{eq_tau0}
\end{equation}
We note that these three dynamical regimes are tightly related to an underlying topological phase transition. As $p$ decreases, the whole system splits
up into a number of smaller subsets or {\em clusters}. Such process displays a critical behaviour near the threshold $p_c\approx 1/N$ \cite{AB01,siam}. Below $p_c$, each node is isolated or linked to a small number of peers. Above $p_c$, a large giant component emerges including a number of nodes of order $O(N)$, whose fraction tends to $1$ as $p\to 1$. This explains the dynamical regimes in eq.(\ref{eq_tau0}).
If $pN\to \infty$ (dense regime), then $\tau\to 0$ and $\rho(x)$ is uniform between $0$ and $1$ as in the initial state, since an infinite number $\langle k_{min}\rangle=pN$ of fitnesses is continuously udpated as on a complete graph. In this case, the step--like behaviour is destroyed. If $pN=d$ with finite $d>1$ (sparse regime), then $\tau$ remains finite as $N\to\infty$, and this is the case considered in ref.\cite{BSrn} that we recover correctly. Finally, if $p$ falls faster than $1/N$ the graph is below the percolation threshold (subcritical regime): the updates cannot propagate and $\tau\to 1$, as for $N$ isolated vertices (only the minimum is continuously updated, which after many timesteps results in pushing all fitness values, except the newly replaced one, towards $1$). 
Therefore the dynamical transition is rooted in an underlying topological phase transition. This previously unrecognized property is fundamental and, as we show below, is also general.

\section*{Fitness--dependent complex networks}
A nontrivial form of $f(x,y)$ must be chosen carefully. 
On static or fitness--independent networks $x_i$ is usually interpreted as the \emph{fitness barrier} against further mutation, and the links are interpreted as feeding relations \cite{BS}. However, once the topology depends on $x$ these two interpretations are difficult to concile. The coupling we have introduced requires consistent interpretations of $x$ and of the links. Also, the form of $f(x,y)$ must be consistent with the feature that the updates of $x$ propagate through the network determined by it. This very instructive aspect must characterize any model with coupled topology and dynamics, and reduces significantly the arbirariness introduced in the static case. 
Here we suggest that the simplest self--consistent interpretation is the following. Since there is no external world in the model, the environment experienced by a species is simply the set of its ecological interactions. Now let $x_i$ represent the fitness (rather than the barrier) of $i$, and let a link between two species mean `being fit to coexist with each other' (i.e. it represents an undirected, non--feeding interaction beneficial to both). The more a species is connected to other species, the more it is fit to the environment. This picture is self--consistent provided that the larger $x$ and $y$, the larger $f(x,y)$. Following the results of refs.\cite{newman_origin,likelihood}, the simplest unbiased \cite{likelihood} choice for such a function is
\begin{equation}
f(x,y)=\frac{zxy}{1+zxy}
\label{fermi}
\end{equation}
where $z$ is a positive parameter controlling the number of links. This choice generates a network with a nonrandom, fitness dependent expected degree sequence \cite{newman_origin,likelihood}, which in this case is not known a priori and will be determined by the fitness distribution at the stationary state. All other higher--order properties are completely random, except for the structural correlations induced by the degree sequence \cite{newman_origin,likelihood}. It therefore represents the fitness--dependent version of the so--called \emph{configuration model} \cite{siam,maslov}. As we show later on, structural correlations have an important impact on the dynamics. 
With the above choice, $\rho(x)$ can be directly computed analytically through eq.(\ref{eq_rho2}). However we write it in a different form, which is equivalent when $N\to\infty$, in order to solve also more complicated integrals involving it later on. We use the trick $\langle f(x,m)\rangle\approx f(x,\langle m\rangle)$ where the angular brackets denote an average over the distribution $q(m)$ of the minimum fitness, that is $\tau^{-1}\int_0^\tau f(x,m)dm\approx (zx\tau/2)/(1+zx\tau/2)$. As we show in a moment, when $N\to\infty$ this approximated expression becomes exact. Then  eq.(\ref{eq_rho2}) yields 
\begin{equation}
\rho(x)=\left\{\begin{array}{ll}
(\tau N)^{-1} &x<\tau\\
(\tau N)^{-1}+2/(zN\tau^2x) &x>\tau
\end{array}\right.
\label{eq_rhofermi}
\end{equation}
where $\tau$ is the solution of eq.(\ref{eq_norm}), which reads 
\begin{equation}
\frac{1}{\tau}+\frac{1}{z\tau^2}\log\frac{1}{\tau^2}=N
\label{eq_sol}
\end{equation}
If $z$ remains finite as $N\to\infty$, or in other words if $zN\to\infty$, then the trivial solution is $\tau\to 0$. On the other hand, we find $\tau\ne0$ if $zN$ remains finite as $N\to\infty$. To obtain the value of $\tau$ in this case, note that for a nonzero solution the term $1/\tau$ in the above expression is finite and negligible for $N$ large enough. Multiplying both sides by $z$ yields 
\begin{equation}
\frac{1}{\tau^2}\log\frac{1}{\tau^2}=zN
\label{eq_sol2}
\end{equation}
whose solution is $\tau=\sqrt{\frac{\phi(zN)}{zN}}$, where $\phi(x)$ is the ProductLog function defined as the solution of $\phi e^\phi=x$. 
Putting these results together, we have
\begin{equation}
\tau=\sqrt{\frac{\phi(zN)}{zN}}\to
\left\{\begin{array}{lll}
1&zN\to 0\\
\sqrt{\phi(d)/d}&zN=d\\
0&zN\to \infty
\end{array}\right.
\label{eq_tau}
\end{equation}

As for random graphs, we find a marked transition as the scaling of $z$ changes from $N^{-1}$ to more rapidly decaying. This suggests an analogous underlying percolation transition. As we show below, this is indeed the case. We can therefore still refer to the subcritical, sparse and dense regimes. 
Note that as $N\to\infty$ we have $f(x,y)= zxy$ in the sparse and subcritical regimes since $zxy<z\ll 1$, which implies that we can neglect $zxy$ in the denominator of eq.(\ref{fermi}). 
Therefore the expression $\langle f(x,m)\rangle = f(x,\langle m\rangle)$ is exact. On the other hand, in the dense regime we have $\tau\to 0$, which again implies the same expression since $q(m)$ becomes the Dirac delta function $\delta(m)$. Therefore our trick to use the above expression turns out to be exact in all regimes for $N\to\infty$.

\section*{Results and discussion}
\begin{figure}
\centerline{\psfig{figure=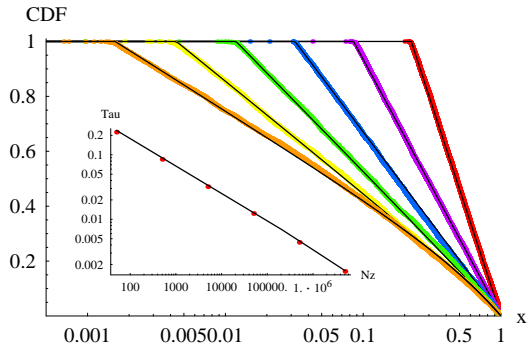,width=8cm} }
\caption{
{\bf Stationary fitness distribution and threshold}. 
Main panel: cumulative density function $\rho_>(x)$ in log--linear axes. From right to left, $z = 0.01$, $z = 0.1$, $z = 1$, $z = 10$, $z = 100$, $z=1000$ ($N=5000$).  Inset: log--log plot of $\tau(zN)$. Solid lines: theoretical curves, points: simulation results.}
\label{fig_px}
\end{figure}

The main panel of fig.\ref{fig_px} shows the cumulative density function (CDF) of the fitness $\rho_>(x)$, while the inset shows a plot of $\tau(zN)$. The theoretical results are in excellent agreement with numerical simulations. 
As predicted by eq.(\ref{eq_rhofermi}), $\rho(x)$ is the superposition of a uniform distribution and a power--law with exponent $-1$. 
For $z\ll 1$ we have $f(x,y)\approx zxy$ and $\rho(x)\propto x^{-1}$ for $x>\tau$. This purely power--law behaviour, that becomes exact in the sparse regime $z=d/N$ for $N\to\infty$, results in a logarithmic CDF looking like a straight line in log--linear axes. 
Note that, despite the value of the exponent, the presence of a nonzero lower threshold ensures that $\rho(x)$ is normalizable. This mechanism may provide a natural explanation for the onset of Pareto distributions with a finite minimum value in real systems.
By contrast, for large $z$ the uniform part is nonvanishing and $\rho(x)$ deviates from the purely power--law behaviour. 
The decay of $\rho(x)$ for $x>\tau$ is a completely novel outcome of the extremal dynamics due to the feedback with the topology: now the fittest species at a given time is also the most likely to be connected to the least fit species and to mutate at the following timestep. Being more connected also means being more subject to changes. This enriches the coexistence patterns displayed on static networks. 

\begin{figure}
\centerline{\psfig{file=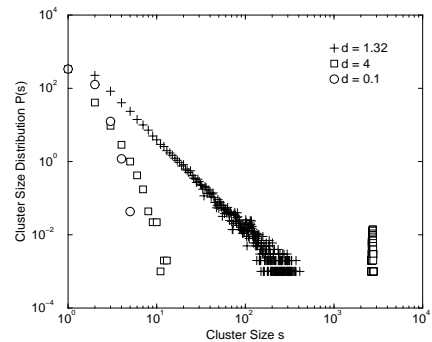,width=5.6cm} }
\caption{
{\bf Cluster size distribution}. Far from the critical threshold 
($d=0.1$ and $d=4$), $P(s)$ is well peaked. At $d_c=1.32$, $P(s)\propto s^{-\gamma}$ with $\gamma=2.45\pm 0.05$. Here $N=3200$.}
\label{cluster_size_distribution}
\end{figure}
\begin{figure}
\centerline{\psfig{file=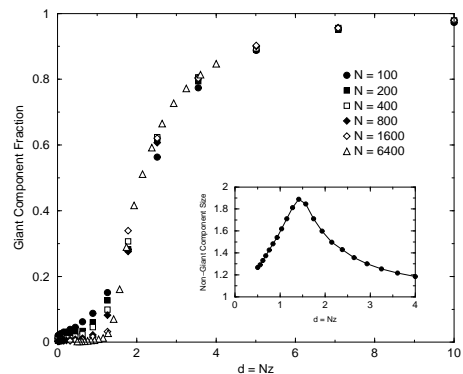,width=6cm} }
\caption{
{\bf Behaviour at the percolation threshold}. Main panel: the fraction of nodes in the giant component for different
network sizes as a function of $d$. Inset: the non-giant
component average size as a function of $d$ for $N=6400$.}
\label{giant_component}
\end{figure}
We now check the conjectured percolation transition. For different system sizes, we find that the cluster size distribution $P(s)$
displays power--law tails when the control parameter $d\equiv zN$ approaches a critical value $d_c =
1.32 \pm 0.05$ (corresponding to $z_c=d_c/N$), suggesting the onset of a second--order
percolation--like phase transition.  As shown in figure \ref{cluster_size_distribution}, $P(s) \propto s^{-\gamma}$ with $\gamma=2.45 \pm 0.05$ at the phase transition.
Fig. \ref{giant_component} shows that the average
fraction of nodes in the largest component remains negligible for
$d<d_c$, whereas it takes increasing finite values above $d_c$. 
As an additional check, following the method adopted in
ref.\cite{newman03}, we have plotted the average size fraction of non--giant
components, which diverges (in the infinite volume limit) when $P(s)$ decays algebraically as reported in the
inset of fig. \ref{giant_component}.

\begin{figure}[h]
\centerline{\psfig{figure=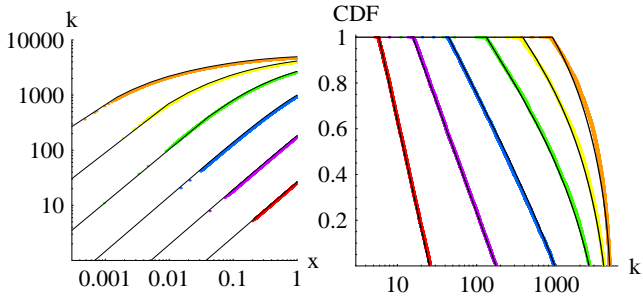,width=9cm} }
\caption{
{\bf Fitness--dependence and cumulative distribution of the degrees}.
Left: $k(x)$ ($N=5000$; from right to left, $z=0.01$, $z=0.1$, $z=1$, $z=10$, $z=100$, $z=1000$). Right: $P_>(k)$ (same parameter values, inverse order from left to right). Solid lines: theoretical curves, points: simulation results.}
\label{fig4}
\end{figure}
Even if one of the most studied properties of the BS model on regular lattices is the statistics of avalanches characterizing the SOC behaviour \cite{BS}, we do not consider it here. This is because, as shown in ref.\cite{notsoc}, when considering long--range \cite{BSrn} instead of nearest--neigbour connections, it can lead to a wrong assessment of the SOC state, which is put into question by the absence of spatial correlations even in the case that avalanches are power--law distributed. 
Rather, we further characterize the topology at the stationary state by 
considering the degree distribution $P(k)$ and the degree correlations. Using eq.(\ref{eq_kx}), we find that the average degree $k(x)$ of a vertex with fitness $x$ is
\begin{equation}
k(x)=\frac{2}{z\tau^2}\ln\frac{1+zx}{1+z\tau x}+\frac{zx-\ln(1+zx)}{z\tau x}
\label{eq_k}
\end{equation}
Similarly, through eq.(\ref{eq_pk}) we can determine the analytical expression for the cumulative degree distribution $P_>(k)$. 
As shown in fig.\ref{fig4}, $k(x)$ is linear for small $z$ since $f(x,z)\approx zxy$, while for large $z$ it saturates to the maximum value $k_{max} = k(1)$.
This implies that in the sparse regime $P(k)$ mimics $\rho(x)$ and is characterized by the threshold value $k(\tau)$ and by a power--law decay $P(k) \propto k^{-1}$ above it (see fig.\ref{fig4}). 
Note that here $\tau$ remains finite even if $P(k)\propto k^{-\gamma}$ with $\gamma<3$, in striking contrast with what obtained on static scale--free networks \cite{BSsf,BSkim,BSkahng}. 
Differently, for large $z$ the saturation $k \rightarrow  
k_{max}$ translates into a cut--off that makes $P(k)$ deviate from the pure power--law behavior for $k>k(\tau)$. 
As shown in refs.\cite{newman_origin,mywtw}, this saturation determines anticorrelation between the degrees of neighboring vertices (disassortativity) and a hierarchy of degree--dependent clustering coefficients as observed in many real--world networks (this is not shown here for brevity). 
As $N\to\infty$, these correlations vanish in the sparse regime ($\tau>0$), while they survive in the dense regime ($\tau\to 0$). Structural correlations and a nonzero threshold $\tau$ are then mutually excluding in this model, which is another interesting effect of the feedback we have introduced. We finally note that with a different choice of $f$, one can have any exponent for the power--law distribution of fitnesses, and therefore for the degree distribution as well. This makes our model completely flexible in order to reproduce any desired topological property.

Our results represent a first step into the unexplored domain of systems with generic self--organized coupling between dynamics and topology. A huge class of  such processes needs to be studied in the future, to further understand the unexpected effects of this coupling.\\

Correspondence and requests for materials should be addressed to G.C. (guido.caldarelli@roma1.infn.it).

G.C. acknowledges D. Donato for helpful discussions. 
This work was partly supported by the European Integrated Project DELIS.

D.G. developed the theory and performed computer simulations, 
A.C. performed computer simulations, 
G.C. planned the study and developed the theory.

The authors declare no competing financial interests.

\end{document}